\documentclass[10pt]{article}
\usepackage[letterpaper]{geometry}
\usepackage{hicss}
\usepackage[T1]{fontenc}
\DeclareFontShape{T1}{cmss}{b}{n}{<->ssub*cmss/bx/n}{}
\usepackage[utf8]{inputenc}
\usepackage{subcaption}
\usepackage{textcomp}
\usepackage{times}
\usepackage[none]{hyphenat}
\usepackage{url}
\usepackage{latexsym}
\usepackage{indentfirst}
\usepackage{graphicx}
\graphicspath{{images/}}
\usepackage[
    style=apa,
  ]{biblatex}
\title{The Social Cost of an AI Teammate: How an Artificial Teammate Reshapes Human-Human Communication in Small-Team Decision-Making}

\author{Nia Nixon \\
 School of Education \& Dept. of Cognitive Science \\
 University of California, Irvine \\
 {\underline{ dowelln@uci.edu}} \\ \\
 Mohammad Amin Samadi \\
 School of Education \\
 University of California, Irvine \\
 {\underline{ msamadi@uci.edu} } \\ \And
 Jaeyoon Choi \\
 School of Education \\
 University of California, Irvine \\
 {\underline{ jaeyoon.choi@uci.edu} } \\ \\
 Seehee Park\\
 School of Education \\
 University of California, Irvine  \\
 {\underline{ seeheep@uci.edu} } \\ \And
 Pedro Martins De Bastos \\
 School of Education \\
 University of California, Irvine  \\
 {\underline{ pedrom4@uci.edu} } \\ \\
 Spencer JaQuay\\
 School of Social Ecology \\
 University of California, Irvine  \\
 {\underline{ sjaquay@uci.edu} } \\ }

\date{}

\begin{document}
\maketitle

\begin{abstract}
Conversational AI is increasingly positioned as a teammate rather than a tool, yet little is known about how its presence reshapes communication among human teammates. We examined sociocognitive communication dynamics in a team decision-making task using Group Communication Analysis (GCA), surveys, and lexical analyses. Sixteen teams of two students and an AI teammate and 17 all-human teams of three completed a high-stakes moral-dilemma task. Mixed-effects models showed that the AI participated more and was more internally cohesive but contributed less new and less dense content than students. Compared with controls, students on AI-inclusive teams showed lower participation, responsivity, and social impact and reported lower belonging, status, and connection. Event-locked models found no condition differences in trajectories of socioemotional content or student–student linguistic style matching. These findings suggest a social cost to students’ experienced team membership and relational integration in AI-inclusive teams, although the temporal analyses do not establish when this cost emerged. We discuss implications for future research in voice-based and longitudinal settings.
\end{abstract}

\par\vspace{1.5\baselineskip}

\section{Introduction}

Conversational AI agents increasingly join teams as participants rather than tools \parencite{Seeber2020}. They take on
knowledge-intensive roles \parencite{Oberhofer2024} and reshape team
communication and coordination \parencite{Zhang2023}.
Accordingly, human-AI teaming research has focused on making AI a trusted,
accepted, and positively regarded teammate \parencite{ONeill2022}. This focus overlooks how AI reshapes relationships
\textit{among humans} and the social fabric of collaborative work.

Most HAT research measures human perceptions of AI---trust,
anthropomorphism, ingroup membership, and satisfaction with its role
\parencite{Oberhofer2024,ONeill2022}. Less work examines the
AI's sociocognitive role in group discourse or its effects on human-human
ties. Two lines of evidence suggest consequences: human-AI teams produce more
task-oriented and fewer socio-emotional messages, and show weaker team
mental-model formation and swift trust under some conditions \parencite{Zhang2023,Oberhofer2024}; in human-robot teams, robot behavior also
ripples into human conversational dynamics \parencite{Sebo2018}. Yet how AI reshapes human relationships and belonging remains
understudied.

To address this gap, we apply Group Communication Analysis (GCA; Dowell et
al., 2019) to a chat-based human-AI team corpus, pairing its measures with
surveys of belonging, status, and felt value and lexical analyses of
socio-emotional content, linguistic style matching, and deliberative stance.
Teams deliberated whether to attempt a mountain rescue under contested risk
and incomplete information, a context in which relational coordination is
consequential. We also test whether human-human communication changes as
teams work through the moral reveal \parencite{Ricca2020}. Here, social cost
means diminished human-human communication and belonging. The study is one
instantiation of a broader program treating AI-teammate design as manipulable
and pairing communication processes with individual differences and lived experience \parencite{choi2026read,cai2024advancing,samadi2024minds}; the platform is described in
Section 3.5.

This paper makes three contributions: (1) characterizing the AI's
sociocognitive role relative to human teammates; (2) measuring AI presence's
social cost for human members; and (3) examining its temporal pattern, finding
that measured indicators remain relatively stable during the observed
interaction.

\section{Background and Research Questions}

\subsection{AI teammates in small groups}

The framing of AI systems as teammates rather than tools has moved from
a research agenda into an active empirical program. The original call by
Seeber and colleagues (2020) mobilized 65 collaboration scientists
around the question of what changes when a machine joins a team,
identifying design areas spanning the machine artifact, the
collaboration itself, and the institutional context. That call has since
produced a wave of empirical and conceptual work establishing that
conversational agents do not just assist humans in teams; they alter how
teams form, perform, and feel \parencite{ONeill2022,Zamecnik2026}.

Recent studies find effects on team cognition, communication, reasoning,
decision-making, participation, and labor division, but results are mixed:
human-agent teams can show weaker ingroup perceptions and mental models yet
more focused discussion under some conditions \parencite{Do2023,Oberhofer2024}.
Design work therefore emphasizes calibrating AI participation, and meta-analytic
evidence cautions that human-AI teams are not automatically
beneficial \parencite{Oberhofer2024,Vaccaro2024}.

This work has centered on human perceptions of AI and task-level performance.
Less attention has been paid to the \textit{human-human ties} within AI teams:
an artificial member does not make the human exchange a dyad, and those
exchanges retain their own structure and consequences. Building on work
mapping sociocognitive dynamics when AI takes leading or following roles \parencite{choi2026read}, we ask whether AI displaces human relational fabric and at
what social cost.

\subsection{Group Communication Analysis}

Group Communication Analysis (GCA; \textcite{Dowell2019}) combines automated
linguistic and sequential-interaction analyses to characterize sociocognitive
roles in multiparty discourse. Its six measures index participation (sharing
information), internal cohesion (semantic continuity with one's prior turns),
responsivity (uptake of teammates' contributions), social impact (others'
uptake of one's contributions), newness (information not yet shared), and
communication density (information per turn). Responsivity and social impact
instantiate uptake; together, the measures
distinguish interactional roles and relate to collaborative problem-solving
outcomes \parencite{Dowell2020}.

\subsection{The relational channel: cohesion, belonging, and disengagement}

Team discourse also enacts a relational fabric: who is recognized, responded
to, free to dissent, and made to feel belonging \parencite{Bales1950,Bonito1997}. This channel matters for functioning;
psychological safety supports learning, voice, and performance \parencite{Edmondson1999}, including in technology-mediated teams \parencite{Seeber2024}. GCA's
responsivity, social impact, and internal cohesion capture this integration:
who builds on whom, whose contributions are taken up, and how discourse
threads together.

The question of how a non-human team member might disturb that relational fabric is open, but precedent exists. In human-robot teams, a robot's social behavior demonstrably ripples onto its human teammates' trust-related behavior and conversational dynamics, including expressions of vulnerability, explanation of failure, and consolation among the humans themselves
\parencite{Sebo2018}. These findings suggest that an artificial teammate can influence human-human relations, although the direction of that influence may depend on how the agent participates. If an AI occupies substantial team airtime without becoming part of the team’s relational exchange, it may reduce opportunities for human teammates to respond to, elaborate on, and support one another. We refer to this process as \textit{relational displacement}: the occupation of communicative space by an AI in ways that leave the human-human ties around it less elaborated. The empirical question is whether and how this displacement appears in team discourse and in human members’ experiences of the collaboration.

We use \textit{social cost} for the resulting human relational consequence:
diminished human-human communication and belonging, status, or connection. It
differs from trust in AI \parencite{Zamecnik2026} and from
social loafing \parencite{Karau1993}, because the focal cost is weaker
human-human responsivity and integration, not less effort or substantive
contribution. Thus, AI presence is the condition, relational displacement the
proposed process, and social cost the outcome assessed through discourse and
team experience.

\subsection{Research questions}

RQ1. What sociocognitive role does an AI teammate enact in a
team's communication relative to its human
teammates?

RQ2. Does the AI's presence change how the
humans communicate with and relate to each other, in discourse (GCA) and
in subjective team experience (belonging, status, felt value)?

RQ3. Does any effect of AI on social cost immediate, or does it emerge dynamically
over the course of the high-stakes discussion?

\section{Method}

\subsection{Design and participants}

The study took place at a large public research university in Fall 2025. We
recruited 80 undergraduates for a between-teams experiment with 33 configured
teams. TRAIL filled available slots through sequential, greedy random
allocation as students entered the virtual environment; students did not
select their team or condition, and assignment used no demographic
stratification. The allocation targeted approximately balanced total
membership: 16 treatment
teams had two student seats and AI while 17 control teams had
three student seats and no AI. Thus, each condition had three teammate seats,
but treatment replaced one human seat with the AI.

The 33 configurations represented 83 nominal student positions (16 $\times$ 2 + 17 $\times$ 3), but only 80 students were recruited. Five recruited students were excluded from the analytic sample because of missing chat data or dropout, leaving 75 analyzed students: 30 in treatment teams and 45 in control teams. The difference between 83 nominal positions and 80 recruits reflects positions that were not filled; the further reduction from 80 recruits to 75 analyzed students reflects the five exclusions. After these exclusions, the analytic roster was uneven: 14 treatment teams contributed two student records and two contributed one, while 12 control teams contributed three student records, four contributed two, and one contributed one. The participant-level GCA roster therefore contains 91 computational rows (75 student nodes plus 16 AI nodes); the AI rows are not recruited participants and are excluded from survey analyses.

The three-agent team structure was chosen to balance internal validity and tractability. Three agents form a small, complete interaction network in which relational coordination can be observed for every pair, while keeping the text-chat record legible and limiting the meeting noise associated with larger teams. Replacing one human seat with the AI nevertheless means that treatment teams contained fewer human students than control teams. We therefore treat human-member group size as a design limitation, retain per-student outcomes as the primary between-condition measures, and use within-treatment comparisons for RQ1, which avoids a between-condition group-size contrast.

The study was approved by the University of California, Irvine IRB (protocol \#6594). Participation was voluntary. Students received class credit regardless of whether they participated, and an alternative class activity was offered for those who opted-out.

\subsection{TRAIL: The Experimental Platform}

We used TRAIL \parencite{samadi2026trail}, a web platform for real-time experiments with human--AI teams. On TRAIL, the AI teammate is a configurable variable: its persona and language model are stored with the experiment and reapplied in every session. The platform also runs group chat, surveys, and random assignment, and it logs every turn. For each batch of messages, a pipeline decides whether the AI should speak and, if so, writes a reply that fits its persona and the team context.

We placed one AI agent, Clever Lamarr, in all 16 treatment teams, with its persona and role held constant. It ran on Gemini 2.5 Flash-Lite. An \texttt{action\_manager} weighed the agent's potential contribution, the group's dynamics, and timing, then returned ``Respond'' or ``Don't respond.'' A \texttt{response\_generator} then wrote the message, speaking as a Gen-Z college peer (medium on 11 personality dimensions, five phases) who avoided assistant-like language and did not rush to solve the problem. Replies were one sentence of 10 words or fewer (50-token cap), at temperature 0.5 instead of the default 1.0. We chose this length to mimic student texting and to leave the work of building a shared problem representation to the students \parencite{Roschelle1995}. The full prompts are available online.\footnote{\url{https://osf.io/v53sx/}.}

These choices may have affected how often the AI spoke and how students took up its turns. Model-specific sycophancy may also shape interaction \parencite{Cheng2026}; we did not measure it in this study, and our single-model design confounds the effects of model, persona, and prompt.

\subsection{Task}

Teams completed a text-only chat-based decision task as a search-and-rescue planning team. A climber was stranded near the summit of a remote ridge during a developing storm, and teams had to recommend whether to attempt a rescue and justify their recommendation to a supervising authority. Midway through the conversation, the chat introduced new information: the climber had ignored posted weather warnings and set out without a registered route. This reveal served as a designed moral perturbation, testing whether autonomous risk-taking should change the team’s obligation to act.

Before this study, run in Fall 2025, the research team ran test sessions with graduate students to test both the task and the AI's behavior. These sessions were used to assess whether the two-phase scenario and instructions elicited meaningful discussion. We administered these pilots to a mixture of graduate and undergraduate students. The task underwent refinements before the Fall 2025 study version was launched, and the team judged the final version of the task suitable. These refinements targeted the instruction wording and contextual phrasing rather than adding or removing substantive scenario content. The scenario supplied enough information for participants to reason about compassion, rescuer safety, uncertainty, responsibility, and precedent; it did not require specialized search-and-rescue knowledge.

\subsection{Measures}

GCA (\textcite{Dowell2019}) was applied to all chat turns from both conditions, treating each participant, including the AI in treatment teams, as a node in the team's discourse network. The six GCA dimensions (participation, internal cohesion, overall responsivity, social impact, newness, and communication density) were computed per participant using the standard sequential-semantic pipeline; turn-window and embedding parameters followed established practice for short-form chat corpora. GCA is an automated computational measure rather than a human-coded content analysis: it is derived from linguistic features and sequential-semantic relations in the chat corpus. Accordingly, there was no human-coding stage and no intercoder-reliability statistic to report for the GCA results.

Team-experience surveys were administered immediately following the
task. Two composite measures derived from this instrument anchor our
subjective outcomes. Belonging was indexed using a 11-item composite adapted from \textcite{Chung2020} assessing whether students felt they were a valued member of, were accepted by, and were connected to their team. Status was indexed using a 12-item composite capturing perceived equality of voice and influence
within the team (adapted from \textcite{Jeitziner2025}). Participants responded to all composite items using a 1-7 Likert scale, and items were averaged to create the belonging and status scores.

We additionally examine three single items at the
construct level: \textit{feeling like a valued member of the team},
\textit{feeling connected to other team members}, and \textit{the perception
that one or two members dominated the collaboration}.

For the temporal analyses reported in Section 4.3, we additionally drew
on lexical content categories from the Linguistic Inquiry and Word Count
dictionary (LIWC-22; \textcite{Boyd2022}). The
per-turn socioemotional content score used in the event-locked
sliding-window analysis was the mean of five LIWC-22 categories: Social,
Affiliation, Prosocial, We, and Positive Tone, each expressed as the
percentage of words in the turn belonging to that category.
Student-student linguistic style matching (Niederhoffer \& Pennebaker,
2002) was computed on the nine standard function-word categories
(personal pronouns, impersonal pronouns, articles, prepositions,
auxiliary verbs, adverbs, conjunctions, negations, and quantifiers),
using the windowed cosine matching protocol described by Pennebaker and
colleagues. For the deliberative-stance analysis we coded each turn for
the presence of possibility modals (could, can, might, may, maybe,
perhaps, possibly, possible), hypothetical modals (would, if, suppose,
imagine, hypothetically), and deontic modals (should, ought, must,
shall, gotta, and forms of have/has/need/needs to and supposed to),
deriving a Markov sequence of stance states per team. All of these
analyses operate on the same chat corpus as GCA; they constitute
complementary probes of the sociocognitive processes that unfold over
the course of the interaction.

Individual gender (Female, Male) was measured as a self-reported variable. The sample is
gender-imbalanced, with 63 female and 12 male students and only three
male students in the treatment condition, which precluded a clean
individual-gender moderation analysis and motivated the female-only
sensitivity analysis we report below.

\subsection{Analysis}

We address the three research questions with linear mixed-effects models that account for students' nesting within teams. For RQ1 and RQ2, we fit one model per raw GCA dimension, \texttt{dimension $\sim$ role + (1 $\mid$ team)}, where role distinguishes the AI, treatment students, and control students. The focal RQ1 contrast is AI minus treatment student; the focal RQ2 contrast is treatment student minus control student. For the four student survey outcomes---belonging, status, felt valued, and felt connected---we fit \texttt{outcome $\sim$ condition + (1 $\mid$ team)} after excluding all AI rows. The belonging and status composites were retained as construct-level measures. Models were estimated by maximum likelihood and report raw-scale fixed-effect estimates, standard errors, 95\% Wald confidence intervals, exact $p$-values, and standardized estimates. We additionally regress per-team mean felt value on per-team AI airtime dominance to examine whether AI dominance covaries with the felt cost.

To assess whether the student-level RQ2 contrasts could be attributable to the gender-composition imbalance, we repeat the condition models among the 63 female students in the analytic sample (27 treatment, 36 control). For GCA, each student's score remains the score computed on the intact AI-inclusive network; only the human rows entering the treatment-versus-control model are filtered. The six GCA and four survey sensitivity tests use separate Benjamini--Hochberg FDR families.

GCA is computed on each complete, observed three-member discourse network: treatment teams comprise two students and the AI, whereas control teams comprise three students. This intact-network specification preserves the multiparty interactional system from which participation, responsivity, social impact, and the other GCA dimensions are derived. Removing the AI would change treatment teams from triads to dyads and would therefore change the relational dynamics being measured rather than isolate the same construct within the same network. We consequently estimate and interpret GCA only for the intact team networks and use the student surveys as complementary evidence about students' experienced belonging, status, value, and connection.

For RQ3, we divide each team's student turns into six non-overlapping, equal-count event bins: three before and three after the moral-information reveal, using the conversation midpoint when no event marker is detected. The first primary outcome is the mean of five LIWC-22 categories within each bin: Social, Affiliation, Prosocial, We, and Positive Tone. The second is student--student linguistic style matching (LSM). For each student in each bin, we calculate a word-count-weighted profile across nine function-word categories---personal pronouns, impersonal pronouns, articles, prepositions, auxiliary verbs, adverbs, conjunctions, negations, and quantifiers---and retain profiles containing at least 15 words. We calculate pairwise LSM between eligible students and average across pairs when a team has more than two eligible students. This yields 121 observed team-bin LSM values across 28 teams; missing bin values are not imputed.

For each primary outcome, we fit \texttt{outcome $\sim$ time $\times$ condition + (1 + time $\mid$ team)}, with time coded $-3$, $-2$, $-1$, $+1$, $+2$, and $+3$. When a random-slope model was singular or produced invalid fixed-effect standard errors, we used a random-intercept model. Models use all available outcome observations. Running-centroid semantic distance is analyzed with the same specification as a secondary outcome. Analyses restricted to teams with a detected event marker assess sensitivity to the midpoint rule.

Because the outcomes are related, we do not use Bonferroni correction \parencite{Perneger1998}. Instead, we foreground estimates and 95\% confidence intervals and control the Benjamini--Hochberg false-discovery rate within conceptually distinct families: the six RQ1 GCA contrasts, six RQ2 GCA contrasts, four survey contrasts, and the two primary RQ3 time-by-condition interactions for socioemotional content and LSM. The female-only gender-balance sensitivity models form separate six-test GCA and four-test survey families, and the marker-only temporal models form a separate sensitivity family. The secondary running-centroid result is reported with its estimate, confidence interval, and exact $p$-value and is not included in the primary RQ3 family.

\section{Results}

\subsection{RQ1: The AI's socio-cognitive role}
The mixed-effects models identify a distinctive GCA profile for the AI relative to its student teammates. The AI had higher participation ($b = 0.237$, $SE = 0.031$, 95\% CI $[0.177, 0.297]$, $q < .001$) and internal cohesion ($b = 0.078$, $SE = 0.021$, 95\% CI $[0.036, 0.119]$, $q < .001$), but lower newness ($b = -0.100$, $SE = 0.022$, 95\% CI $[-0.144, -0.056]$, $q < .001$) and communication density ($b = -0.180$, $SE = 0.042$, 95\% CI $[-0.262, -0.098]$, $q < .001$). Differences in overall responsivity ($b = -0.001$, $SE = 0.008$, 95\% CI $[-0.017, 0.014]$, $q = .870$) and social impact ($b = -0.008$, $SE = 0.007$, 95\% CI $[-0.021, 0.005]$, $q = .254$) were small and imprecise. Thus, the AI was a high-volume, self-cohesive, content-light contributor, without a clear difference from treatment students in relational uptake.

\begin{table*}[t]
    \centering
    \small
    \caption{Mixed-effects contrasts between the AI and treatment students across GCA dimensions. Positive estimates indicate higher scores for the AI.}
    \label{tab:rq1-mixed-effects}
    \begin{tabular}{@{}lrrrr@{}}
    \hline
    GCA dimension & Estimate & SE & 95\% CI & FDR $q$ \\
    \hline
    Participation & $0.237$ & $0.031$ & $[0.177, 0.297]$ & $<.001$ \\
    Internal cohesion & $0.078$ & $0.021$ & $[0.036, 0.119]$ & $<.001$ \\
    Overall responsivity & $-0.001$ & $0.008$ & $[-0.017, 0.014]$ & $.870$ \\
    Social impact & $-0.008$ & $0.007$ & $[-0.021, 0.005]$ & $.254$ \\
    Newness & $-0.100$ & $0.022$ & $[-0.144, -0.056]$ & $<.001$ \\
    Communication density & $-0.180$ & $0.042$ & $[-0.262, -0.098]$ & $<.001$ \\
    \hline
    \end{tabular}
    \end{table*}

\begin{figure*}[thb]
\centering
\includegraphics[width=0.50\textwidth]{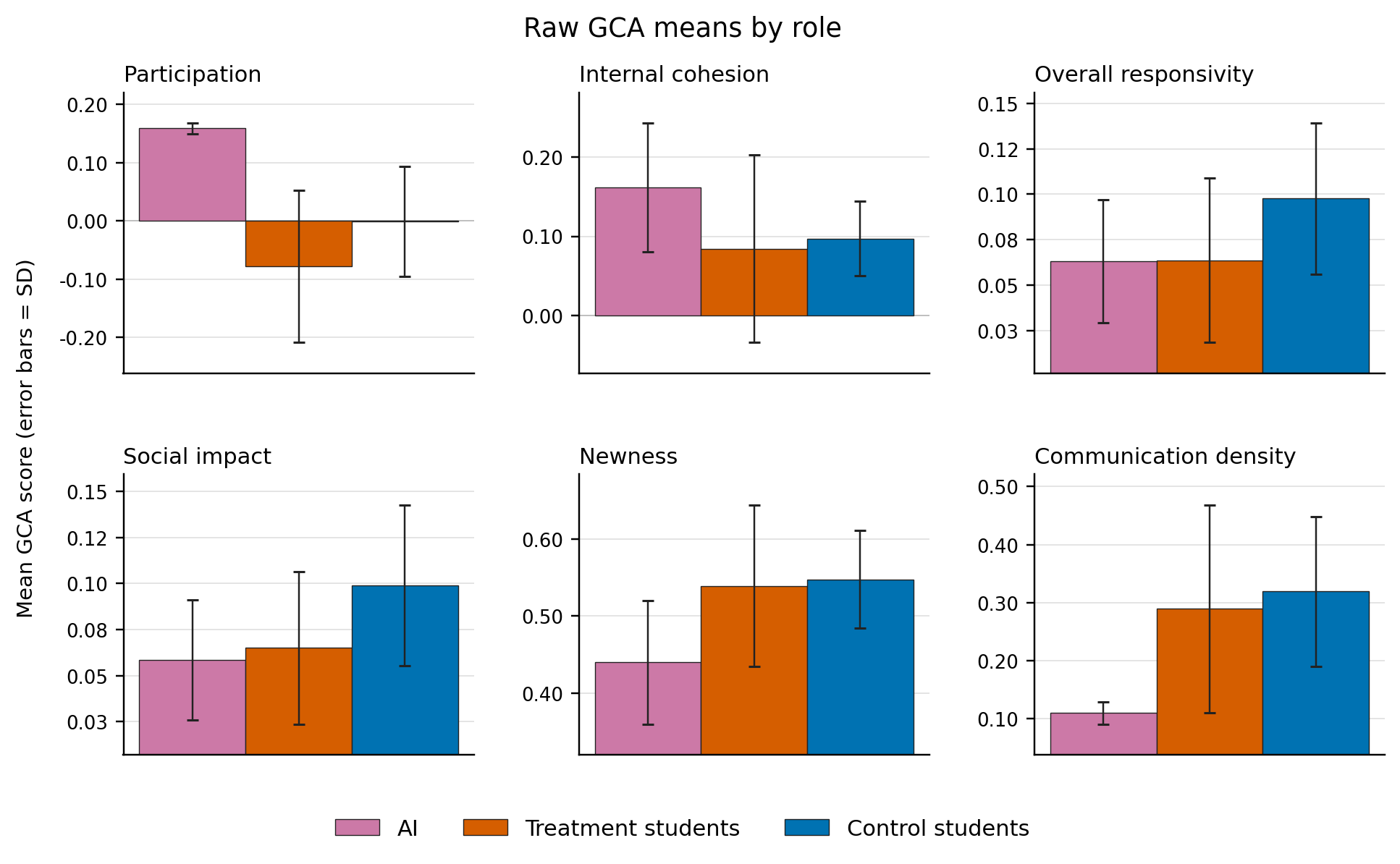}
\caption{Raw GCA means by role across the six dimensions. Bars show the AI teammate, treatment students, and control students; error bars are within-role standard deviations. Each panel uses the raw scale for that dimension, so comparisons are within dimension rather than across dimensions. Counts are based on nonmissing GCA scores and may vary by dimension.}
\label{fig:gca-profile}
\end{figure*}

The pattern matters for what it does not show. The AI talks the most,
but its contributions are not preferentially taken up by its student
teammates: it is not relationally privileged in the group. The cost, as
the next section shows, is borne not by the AI but by the humans.

\begin{figure*}[t]
\centering

\begin{subfigure}[t]{0.5\textwidth}
    \centering
    \includegraphics[width=\linewidth]{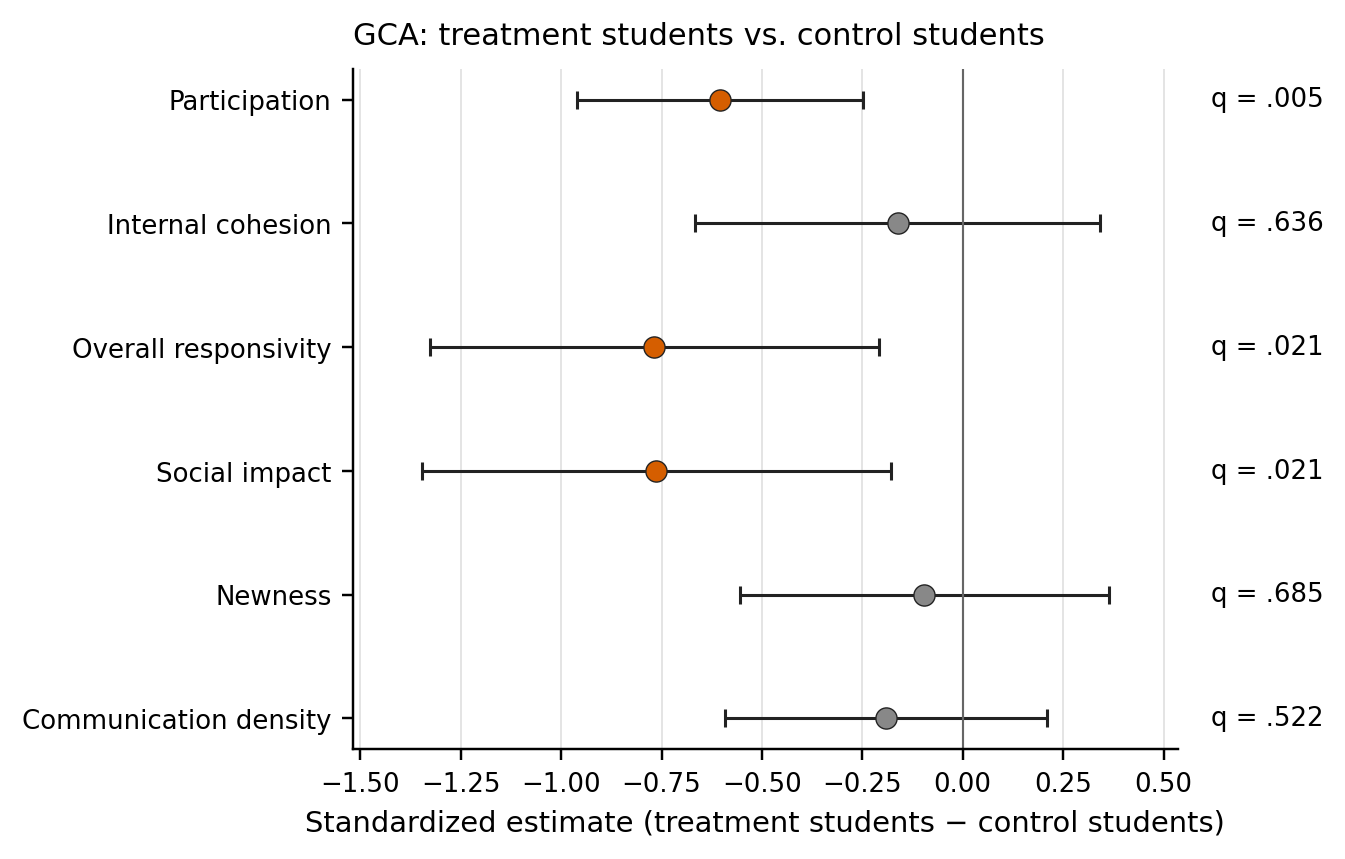}
    \caption{RQ2 GCA comparison of treatment and control students. Points show standardized fixed-effect estimates with 95\% Wald confidence intervals.}
    \label{fig:student-gca-effects}
\end{subfigure}
\hfill
\begin{subfigure}[t]{0.45\textwidth}
    \centering
    \includegraphics[width=\linewidth]{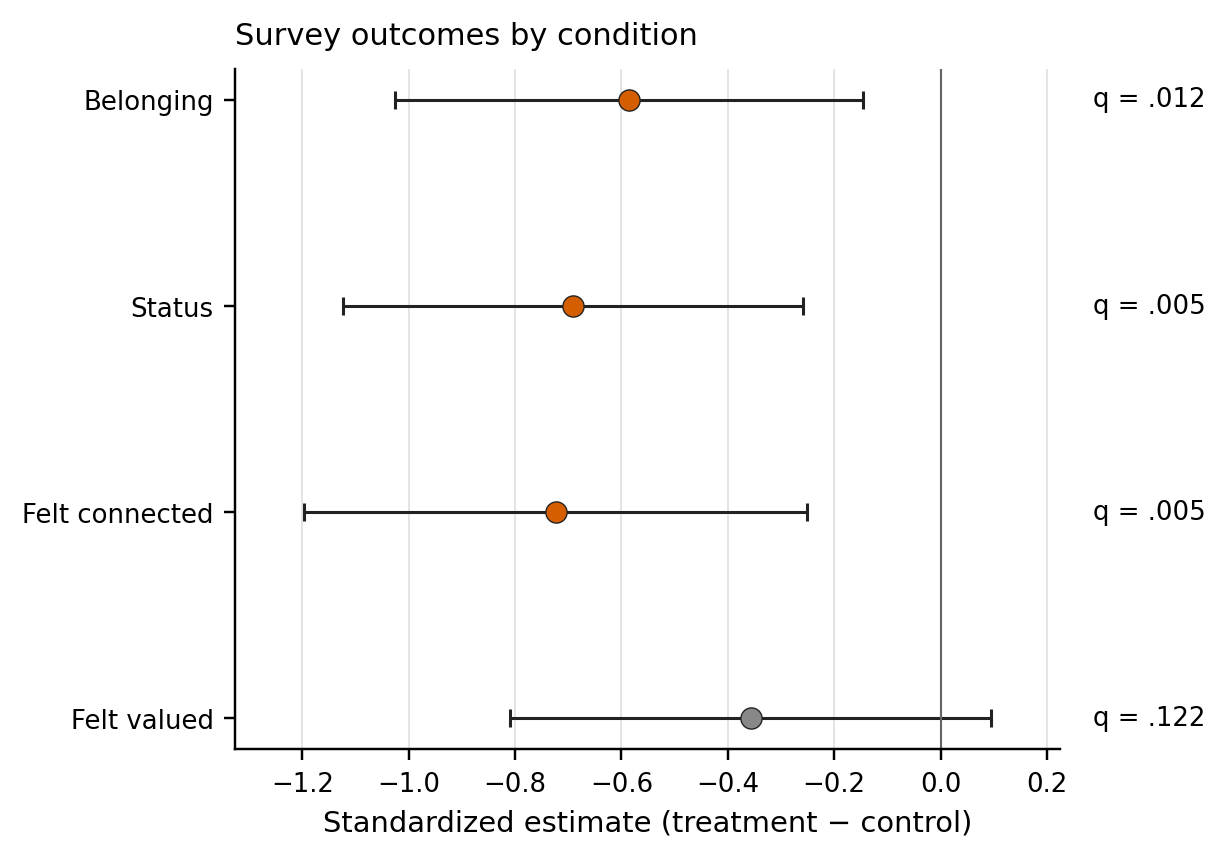}
    \caption{RQ2 student survey outcomes by condition. Points show standardized fixed-effect estimates with 95\% Wald confidence intervals.}
    \label{fig:felt-social-cost}
\end{subfigure}

\caption{RQ2 results for communication and team experience. Orange points have $q < .05$; gray points do not. Negative estimates indicate lower treatment-student scores. Q-values are Benjamini--Hochberg adjusted within the relevant outcome family.}
\label{fig:rq2-results}
\end{figure*}

\subsection{RQ2: The social cost to the humans}

\subsubsection{Discourse disengagement}

In the mixed-effects models predicting GCA dimensions, we found that students in AI teams communicate less with one another than students in
all-human teams. Treatment students had lower participation ($b = -0.077$, $SE = 0.023$, 95\% CI $[-0.123, -0.032]$, $q = .005$), overall responsivity ($b = -0.034$, $SE = 0.013$, 95\% CI $[-0.059, -0.009]$, $q = .021$), and social impact ($b = -0.034$, $SE = 0.013$, 95\% CI $[-0.060, -0.008]$, $q = .021$) than control students. Internal cohesion, newness, and communication density did not differ after FDR adjustment.

The location of these differences is informative. The effects appear on the dimensions that index relational integration, namely how often students take up one another's contributions and how often others take up theirs, rather than on the content-side dimensions of newness and density. Treatment students continue to introduce as much new and as dense content as control students; what they do less of is responding to and being responded to by each other. The AI's presence does not depress the substance of human contribution. It depresses the human-to-human exchange around it.

\subsubsection{Felt cost}

The survey models provide complementary evidence that does not depend on the relational computation of GCA. Relative to control students, treatment students reported lower belonging ($b = -0.695$, $SE = 0.266$, 95\% CI $[-1.217, -0.174]$, $q = .012$), status ($b = -0.332$, $SE = 0.106$, 95\% CI $[-0.540, -0.124]$, $q = .005$), and felt connection ($b = -1.121$, $SE = 0.374$, 95\% CI $[-1.855, -0.388]$, $q = .005$). Felt value was also lower, but the estimate was imprecise and did not survive FDR adjustment ($b = -0.444$, $SE = 0.287$, 95\% CI $[-1.007, 0.118]$, $q = .122$). Together, the GCA and survey findings indicate lower relational integration in AI-inclusive discourse and a felt social cost among students in AI teams.

The female-only sensitivity models preserved this pattern. Treatment-minus-control estimates remained negative for participation ($b = -0.081$, $q = .025$), overall responsivity ($b = -0.034$, $q = .025$), and social impact ($b = -0.035$, $q = .025$) in the intact-network GCA scores, and for belonging ($b = -0.760$, $q = .013$), status ($b = -0.357$, $q = .011$), and felt connection ($b = -1.204$, $q = .006$) in the survey outcomes. Felt value remained negative but imprecise ($b = -0.380$, $q = .225$). These estimates are reported as a gender-balance sensitivity analysis, not as a replacement for the full-sample models.

The felt cost also scales with the discourse-level cost. Across the 16
treatment teams, the more the AI dominated the team's
airtime, the less valued students felt (r = -0.52, p = .04). This
relationship was driven entirely by treatment teams; the
within-treatment correlation is the substantive effect, while the
corresponding test in control teams (where there is no AI to dominate)
was not significant. Figure 1 (right) presents both findings: the
condition-level shift in belonging and status, and the within-treatment
team-level scaling of AI airtime dominance with felt value.

\subsection{RQ3: No evidence that the social cost emerged during discussion}

The socioemotional model used 192 observations from 32 teams, and the LSM model used 121 observed bins from 28 teams. The time-by-condition interaction was not detected for either LIWC socioemotional content ($b = 0.078$, $SE = 0.244$, 95\% CI $[-0.400, 0.557]$, $q = .748$) or student--student LSM ($b = 0.0060$, $SE = 0.0107$, 95\% CI $[-0.0150, 0.0270]$, $q = .748$). Marker-only sensitivity estimates were also imprecise for socioemotional content ($b = -0.293$, $SE = 0.330$, 95\% CI $[-0.939, 0.353]$, $q = .731$) and LSM ($b = 0.0137$, $SE = 0.0397$, 95\% CI $[-0.0641, 0.0915]$, $q = .731$). The secondary running-centroid interaction was close to zero ($b = -0.00039$, $SE = 0.00407$, 95\% CI $[-0.00836, 0.00758]$, $p = .924$). Model-implied changes from $P$-3 to $S$+3 were negative in both conditions for LIWC (control $-2.61$; treatment $-2.14$) and smaller for LSM (control $-0.062$; treatment $-0.026$); these are secondary descriptions of temporal drift. The treatment-control gap changed by $+0.47$ LIWC units (95\% CI $[-2.40, 3.34]$, $q = .748$) and $+0.036$ LSM units (95\% CI $[-0.090, 0.162]$, $q = .748$), providing no evidence that the social-cost gap changed from the earliest to the latest event bin. These results show no detectable change in the measured social cost pattern over the course of the discussion. However, because the study did not include a pre-task baseline, the results do not establish that the effect was present from the beginning. Rather, the findings imply that this may have been an immediate product of the difference between treatment and control conditions.

\section{Discussion}

\subsection{An AI teammate as a relational displacer}

Across the GCA and survey analyses, the AI teammate showed a consistent profile: it participated more, threaded its own contributions tightly across turns, and contributed less new and less dense content than its student teammates. Within the intact team networks, treatment students also showed lower participation, responsivity, and social impact than control students, while reporting lower belonging, status, and felt connection. We interpret this pattern as lower relational integration within AI-inclusive team discourse accompanied by a felt social cost among students. 
Restricting the student-level RQ2 models to female students yielded the same negative direction and FDR-adjusted contrasts for participation, responsivity, social impact, belonging, status, and felt connection; the felt-value estimate remained negative but imprecise. This sensitivity does not eliminate other sources of team-composition variation, but it suggests that the observed RQ2 pattern is not explained solely by the gender imbalance.

We interpret this pattern as consistent with \textit{relational displacement}. By occupying communicative bandwidth without prompting equivalent human-human elaboration, the AI appears to reduce relational exchange among students rather than simply add another source of task input. This pattern bears some resemblance to \textit{production blocking}, in which the structure of group interaction interferes with members' ability to express and develop ideas \parencite{Nijstad2003}. The concepts are related but distinct: production blocking concerns constraints on idea generation, whereas relational displacement concerns the weakening of human-human uptake and relational integration when an AI occupies a central position in the conversation. 

Relational displacement therefore provides a way to understand how an AI teammate may generate social cost. Whereas relational displacement describes the interactional process, social cost refers to its consequences for human teammates, including lower human-human responsivity and weaker experiences of belonging, status, and connection. The present findings suggest that the AI’s high-volume, content-light participation may have altered the relational channel among students without simply reducing their effort or substantive contribution.

Overall, these findings extend human-AI teaming research by shifting attention from how humans relate to AI toward how AI participation reshapes relationships among human teammates. They suggest that the consequences of introducing an AI teammate cannot be evaluated solely through trust, acceptance, task performance; relational integration among the human members is also an important outcome.

This perspective also points to a broader design consideration. AI teammates should be configured not only to contribute useful task information but also to preserve opportunities for human-human exchange. Calibrated participation, reduced airtime dominance, and prompts that encourage human teammates to respond to and build on one another’s contributions may help sustain the team’s relational channel.

\subsection{No detectable change in social cost indicators over time}
The temporal analyses did not detect different within-session trajectories between treatment and control teams. Across six non-overlapping event bins, the linear time-by-condition interactions for socioemotional content and student--student LSM were not retained after FDR adjustment; the secondary running-centroid estimate was likewise close to zero and imprecise. Our estimates for the change in LIWC and LSM metrics were also not significant between the earliest and latest time bins. Taken together, these results suggest that the measured social-cost indicators remained relatively stable across the observed interaction. Because the study did not include a pre-task baseline or a prespecified early-period comparison, the temporal analyses speak to the stability of the pattern during the discussion rather than the precise point at which it first emerged.

Several limitations qualify the present findings. The sample of 33 teams limits power, especially for moderation and temporal interactions, and several mixed models estimated team variances near zero or reached boundary solutions; we therefore foreground estimates and confidence intervals and interpret nulls cautiously. GCA captures each participant's position within the intact multiparty discourse but does not distinguish whether students' uptake was directed toward the AI or toward other human teammates. The study also used a single AI persona, a text-based interface, a single-session design, and a morally framed decision task, so the findings may differ with other designs, modalities, time scales, or collaborative contexts.

\subsection{Open questions for human-AI teaming}

What questions does the construct of social cost open for human-AI teaming? The present study shows a condition difference in human-human communication and team experience that remained relatively stable across the observed interaction. The temporal analyses therefore point to the persistence of the social cost pattern during the discussion, while leaving open the question of when the difference first emerged. Specifically, five follow-on questions structure what comes next.

Does the cost attenuate with habituation, or compound with repetition? A pattern that remains stable within one interaction may still change across repeated interactions, making longitudinal research necessary \parencite{Duan2024}.

Does the cost persist in modalities richer than text? Voice and embodied interfaces may create additional channels for human-human connection and reduce competition for conversational space.

Which AI participation strategies can reduce the cost without sacrificing useful task contributions? Future work should examine calibrated turn-taking, intervention timing, and prompts that encourage human-human uptake.

How does the cost interact with team-member identity? Gender, role, expertise, and team status may shape who responds to whom and who bears the cost in AI-inclusive teams.

How does trust in the AI interact with social cost? Trust may reduce disengagement by enabling comfortable delegation, or amplify it by encouraging students to cede relational attention to the AI.

These five questions are not intended to constitute a comprehensive agenda for human-AI teaming research. Rather, they are the questions the construct of social cost makes legible, each anchoring the design conversation to a measurable effect on the humans rather than to a feature of the AI. If machines as teammates is to be a productive frame, the test of that frame is not only what the machines do, but what they do to the people they are teamed with.

\section*{Acknowledgments}

This research was supported in part by the Jacobs Foundation, grant number 2024-1533-00, Spencer Vision Grant number, 202600129 and the Bill \& Melinda Gates Foundation, grant number INV-068818.

\printbibliography

\end{document}